\documentclass[prb,twocolumn]{revtex4}%

\usepackage{graphicx}
\usepackage{amsmath}
\usepackage{ulem}
\usepackage{color}

\newcommand{\be}{\begin{equation}}
\newcommand{\ee}{\end{equation}}
\newcommand{\bea}{\begin{eqnarray*}}
\newcommand{\eea}{\end{eqnarray*}}
\newcommand{\bean}{\begin{eqnarray}}
\newcommand{\eean}{\end{eqnarray}}

\begin{document}

\draft
\title
{\bf Thermoelectric properties of finite two dimensional
triangular lattices coupled to electrodes}

\author{David M T Kuo }
%and Yia-Chung Chang$^{2}$
%\author{David M T Kuo}
\address{Department of Electrical Engineering and Department of Physics, National Central
University, Chungli, 320 Taiwan}

%\address{$^{2}$Research Center for Applied Sciences, Academic Sinica,
%Taipei, 11529 Taiwan}

%\affiliation{$^3$ Department of Physics, National Cheng Kung
%University, Tainan, 701 Taiwan}

\date{\today}

\begin{abstract}
Novel intrinsic two-dimensional materials have attracted many
researchers' attention. The unusual transport and optical
properties of these materials originate mainly from triangular
lattices (TLs). Therefore, the application of energy harvesting
calls for a study of the thermoelectric properties of 2D TLs
coupled to electrodes. The transmission coefficient of 2D TLs is
calculated by using the Green's function technique to treat
ballistic transports. Especially important among our findings is
the electron-hole asymmetric behavior of the power factor ($PF$).
Specifically, the maximum $PF$ of electrons is significantly
larger than that of holes. At room temperature, the maximum $PF$
of electrons is dictated by the position of the chemical potential
of electrodes near the band edge of TLs. The enhancement of $PF$
with increasing electronic states results from the enhancement of
electrical conductance and constant Seebeck coefficient. When the
band gap is ten times larger than the thermal energy, it is
appropriate to make one-band model predictions for thermoelectric
optimization.
\end{abstract}

\maketitle

\section{Introduction}

Designing a thermoelectric material with a high figure of merit
($ZT$) and optimized power output has been under intensive pursuit
in energy harvesting applications
[\onlinecite{Mahan}-\onlinecite{ChenG}].The dimensionless figure
of merit $ZT=S^2G_eT/\kappa$ depend on the Seeback coefficient
($S$), electrical conductance ($G_e$) and thermal conductance
($\kappa$) of the material. Thermoelectric materials with a
delta-function transmission coefficient show impressive $ZT$
values[\onlinecite{Mahan}]. This inspires one to study the
thermoelectric properties of individual quantum dot (QD)
systems[\onlinecite{MahanG},\onlinecite{Kuo1}-\onlinecite{Kuo2}].
However, such QD systems yield low electrical power outputs.
Another approach to increasing $ZT$ is to reduce the thermal
conductance of thermoelectric
materials[\onlinecite{Chen}],[\onlinecite{Venkatasubramanian}]
-\onlinecite{Harman}]. Although the ZT value of 3D QD
superlattices can reach a remarkable value of
two[\onlinecite{Harman}], it is difficult to raise it to larger
than three. Two dimensional systems offer high potential to
achieve $ZT>3$ due to their reduced phonon thermal conductance
($\kappa_{ph}$).[\onlinecite{Chen}] If a material is found with
$ZT \ge 3$, thermoelectric refrigerators will become competitive
against conventional compressor-based systems[\onlinecite{Kagan}].
Moreover, QD superlattice systems suffer from the problem of size
and position fluctuation, which seriously reduces the $G_e$ and
$ZT$ of QD arrays[\onlinecite{Kuo3}]. Solving these problems is
critical for the applications of thermoelectric devices consisting
of QD solid crystals[\onlinecite{Kagan}].

Intrinsic two-dimensional materials (ITDMs) such as 2D transition
metal dichalcogenides (TMDCs) and oxides (TMOs) have attracted
much attention due to their stable structures and widely tunable
electronic band structures by using external forces such as
electric field and strain
[\onlinecite{GeimAK}-\onlinecite{Desai}]. In addition, the thermal
conductivity of ITDMs is much smaller than that of their
corresponding bulk materials
[\onlinecite{Zhao}-\onlinecite{ChangC}]. Although the
thermoelectric properties of ITDMs have been theoretically studied
using the first-principle method
[\onlinecite{Fan}-\onlinecite{Moon}], the contact and size effects
on the electron transport of their heterostructures are still
unclear [\onlinecite{Wang}]. Because the unusual transport and
optical properties of ITDMs originate mainly from the triangular
(hexagonal) lattices (TLs) [\onlinecite{Qian}-\onlinecite{LuAY}],
their potential application to micro power generators and
refrigerators calls for an investigation of the thermoelectric
properties of finite 2D TLs coupled to electrodes as shown in Fig.
1. A finite 2D TMDC also plays an important role in building
quantum registers and nanoscale transistors
[\onlinecite{Li}-\onlinecite{LiJ}]. This study has several
important findings: (a) the power factor ($PF$) shows an
electron-hole asymmetric behavior, (b) the maximum $PF$ of the
electrons is significantly larger than that of the holes, and (c)
the maximum $PF$ of electrons at room temperature occurs at the
position of the chemical potential of electrodes near the band
edge of TLs. Moreover, we also reveal the contact, size, and
geometry effects on the thermoelectric properties of TLs in the
ballistic transport process. These conclusions are meaningful
insights for further improvement of the performance of micro
thermoelectric devices consisting of intrinsic 2D materials.

\begin{figure}[h]
\centering
\includegraphics[trim=2.5cm 0cm 2.5cm 0cm,clip,angle=-90,scale=0.3]{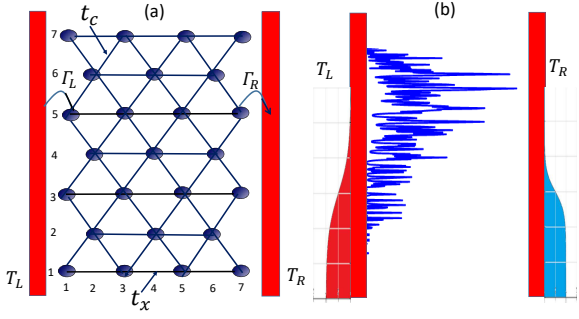}
\caption{(a) Schematic diagram of finite two dimensional
triangular lattices (2D TLs) coupled to electrodes. $\Gamma_{L}$
($\Gamma_R$) denotes the tunneling rate of the electrons between
the left (right) electrode and the leftmost (rightmost) lattices.
$t_x$ and $t_c$ are the nearest neighbor hopping parameters along
the x and $\pm \pi/3$, respectively. (b) Transmission spectra of
finite TLs coupled to electrodes with different equilibrium
temperatures ($T_L$ and $T_R$). }
\end{figure}

\section{Formalism}

To model the thermoelectric properties of a finite 2D TL connected
to the electrodes, the Hamiltonian of the system shown in Fig. 1
is given by $H=H_0+H_{QD}$,[\onlinecite{Haug}] where
\begin{small}
\begin{eqnarray}
H_0& = &\sum_{k,\sigma} \epsilon_k
a^{\dagger}_{k,\sigma}a_{k,\sigma}+ \sum_{k,\sigma} \epsilon_k
b^{\dagger}_{k,\sigma}b_{k,\sigma}\\ \nonumber
&+&\sum_{\ell}^{N_y}\sum_{k,\sigma}
V^L_{k,\ell,j}d^{\dagger}_{\ell,j,\sigma}a_{k,\sigma}
+\sum_{\ell}^{N_y}\sum_{k,\sigma}V^R_{k,\ell,j}d^{\dagger}_{\ell,j,\sigma}b_{k,\sigma}+H.c.
\end{eqnarray}
\end{small}
The first two terms of Eq.~(1) describe the free electron gas in
the left and right electrodes. $a^{\dagger}_{k,\sigma}$
($b^{\dagger}_{k,\sigma}$) creates  an electron of momentum $k$
and spin $\sigma$ with energy $\epsilon_k$ in the left (right)
electrode. $V^L_{k,\ell,j}$ ($V^R_{k,\ell,j}$) describes the
coupling between the left (right) lead with its adjacent lattice
in the $\ell$th row, which counts only for odd rows.
\begin{small}
\begin{eqnarray}
H_{QD}&= &\sum_{\ell,j,\sigma} E_{\ell,j}
d^{\dagger}_{\ell,j,\sigma}d_{\ell,j,\sigma}\\ \nonumber&+&
\sum_{\sigma}\sum_{\ell 1,\ell 2}^{N_y}\sum_{j1,j2}^{N_x} t_{\ell
1,\ell 2, j1, j2} d^{\dagger}_{\ell 1,j1,\sigma} d_{\ell
2,j2,\sigma}+H.c,
\end{eqnarray}
\end{small}
\begin{equation}
t_{\ell 1,\ell 2,j1, j2}= \{ \begin{array}{ll} -t_{c} &
if~|\ell 1-\ell 2|=1, |j1-j2|=1  \\
-t_{x} & if~|\ell 1-\ell 2|=0, |j1-j2|=2
\end{array},
\end{equation}
where { $E_{\ell,j}$} is the atomic energy level of lattice in the
TLs. For simplicity, we have considered one orbit for each atom.
The spin-independent $t_{\ell 1, \ell 2, j1, j2}$ describes the
electron hopping strength between the lattices, which follows the
regulation of Eq. (3). $d^{\dagger}_{\ell 1,j1,\sigma} (d_{\ell
2,j2,\sigma})$ creates (destroys) one electron in the lattice at
the $\ell$th row and $j$th column.

To study the transport properties of a finite TL junction
connected to electrodes, it is convenient to use the
Keldysh-Green's function technique[\onlinecite{Haug}]. Electron
and heat currents leaving electrodes can be expressed as
\begin{eqnarray}
J &=&\frac{2e}{h}\int {d\varepsilon}~
T_{LR}(\varepsilon)[f_L(\varepsilon)-f_R(\varepsilon)],
\end{eqnarray}
and
\begin{eqnarray}
& &Q_{e,L(R)}\\ &=&\frac{\pm 2}{h}\int {d\varepsilon}~
T_{LR}(\varepsilon)(\varepsilon-\mu_{L(R)})[f_L(\varepsilon)-f_R(\varepsilon)]\nonumber
\end{eqnarray}
where
$f_{\alpha}(\varepsilon)=1/\{\exp[(\varepsilon-\mu_{\alpha})/k_BT_{\alpha}]+1\}$
denotes the Fermi distribution function for the $\alpha$-th
electrode, where $\mu_\alpha$  and $T_{\alpha}$ are the chemical
potential and the temperature of the $\alpha$ electrode. $e$, $h$,
and $k_B$ denote the electron charge, the Planck's constant, and
the Boltzmann constant, in that order. $T_{LR}(\varepsilon)$
denotes the transmission coefficient of a finite TL connected to
electrodes, which can be solved by the formula $
T_{LR}(\varepsilon)=4Tr[\hat{\Gamma}_{L}\hat{G}^{r}_{D,A}(\varepsilon)\hat{\Gamma}_{R}\hat{G}^{a}_{D,A}(\varepsilon)]$,
where the matrix of tunneling rates ($\hat{\Gamma}_L$ and
$\hat{\Gamma}_R$) and Green's functions
($\hat{G}^{r}_{D,A}(\varepsilon)$ and
$\hat{G}^{a}_{D,A}(\varepsilon)$) are constructed by fortran
coding. Note that tunneling rates
($\Gamma_{L(R),\ell,j}(\varepsilon)=2\pi\sum_{k}
|V^{L(R)}_{k,\ell,j}|^2
\delta(\varepsilon-\varepsilon_k)=\Gamma_{t,\ell,j}$) are assumed
as energy-independent physical parameters for simplicity's
sake.[\onlinecite{Kuo4}]

The electrical conductance ($G_e$), Seebeck coefficient ($S$) and
electron thermal conductance ($\kappa_e$) can be evaluated by
using Eqs. (4) and (5) with a small applied bias $\Delta
V=(\mu_L-\mu_R)/e$ and cross-junction temperature difference
$\Delta T=T_L-T_R$. We arrived at these thermoelectric
coefficients: $G_e=e^2{\cal L}_{0}$, $S=-{\cal L}_{1}/(eT{\cal
L}_{0})$ and $\kappa_e=\frac{1}{T}({\cal L}_2-{\cal L}^2_1/{\cal
L}_0)$. ${\cal L}_n$ is given by
\begin{equation}
{\cal L}_n=\frac{2}{h}\int d\varepsilon~
T_{LR}(\varepsilon)(\varepsilon-\mu)^n(-\frac{\partial
f(\varepsilon)}{\partial \varepsilon}),
\end{equation}
where $f(\varepsilon)=1/(exp^{(\varepsilon-\mu)/k_BT}+1)$ is the
Fermi distribution function of electrodes at equilibrium
temperature $T$ and chemical potential $\mu$.

\section{ Results and discussion}
Although the density of states (DOS) of triangular lattices has
been studied since very early on, there is hardly any literature
about the calculation of transmission coefficients of a
small-scale 2D TL [\onlinecite{Horiguchi}]. Using the Green's
function techniques, we calculate $T_{LR}(\varepsilon)$ in Fig. 2
as functions of $\varepsilon$ for three  configurations (different
sets of $t_x$ and $t_c$) at
$\Gamma_{L(R),\ell,j}=\Gamma_t=1\Gamma_0$ and $N_x=N_y=N=19$. The
inhomogenous spectra of $T_{LR}(\varepsilon)$ show not only the
distribution of electronic states but also the probability of the
electrons in the electrodes tunneling through these states. The
density of electronic states for a negative regime
($\varepsilon-E_0 <0$) is higher than that of a positive regime
($\varepsilon-E_0>0$). This is consistent with the DOS of
TL.[\onlinecite{Horiguchi}] The distribution range of electronic
states can be explained by an anisotropic electron dispersion
relation $E(\varepsilon)=E_0-2t_x (cos(k_x)+2\gamma cos(k_x/2)
cos(\sqrt{3}k_y/2))$, where $\gamma=t_c/t_x$ and $E_0$ denotes the
atomic energy level. $k_x$ and $k_y$ are dimensionless wave
numbers, which depend on $N_x$ and $N_y$. The lower band edge
(LBE) and upper band edge (UBE) are, respectively,
$-2t_x(1+\gamma^2/2)$ and $2t_x(1+2\gamma)$ when $\gamma=t_c/t_x
\le 2$. LBE is replaced by $-2t_x(2\gamma-1)$ as $\gamma > 2$. For
example, we have $LBE=-36~\Gamma_0$, $UBE=72~\Gamma_0$ and band
width $BW=108~\Gamma_0$ in Fig. 2(a). All physical parameters are
in units of $\Gamma_0$. According to Ref.[\onlinecite{Mahan}],
highly efficient thermoelectric materials prefer
$T_{LR}(\varepsilon)$ with a delta function distribution (or
$\frac{BW}{k_BT} \le 1$). This study will focus on the situation
of $\frac{BW}{k_BT} \gg 1$, because this is an essential condition
of TMDC materials at room temperature.

\begin{figure}[h]
\centering
\includegraphics[angle=-90,scale=0.3]{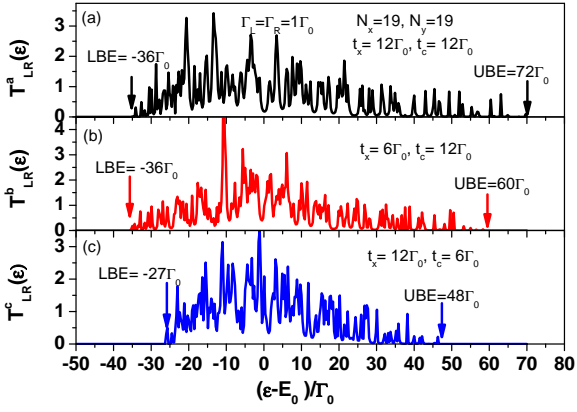}
\caption{Transmission coefficient $T_{LR}(\varepsilon)$ as
functions of $\varepsilon$ for different sets of $t_x$ and $t_c$
at $\Gamma_{L(R),\ell,j}=\Gamma_t=1~\Gamma_0$ and $N_x=N_y=N=19$.
Diagrams (a), (b) and (c) correspond, respectively, to
configurations described by $t_x=12\Gamma_0$ ($t_c=12\Gamma_0$),
$t_x=6\Gamma_0$ ($t_c=12\Gamma_0$), and $t_x=12\Gamma_0$
($t_c=6\Gamma_0$). We set $E_0=0$ throughout this article.}
\end{figure}

Next, we examine the thermoelectric properties of TLs with the
three aforementioned configurations. In Fig. 3 we calculate $G_e$,
$S$ and power factor ($PF=S^2G_e$) as functions of chemical
potential ($\mu$) for different $T_{LR}(\varepsilon)$
configurations at $k_BT=2.5\Gamma_0$ and $k_BT=5\Gamma_0$,
respectively. Each curve of $G_e$ at a finite temperature has two
components resulting from the resonant tunneling procedure (RTP)
and thermionic-assisted tunneling procedure (TATP), respectively
(see Fig. 6(d)). When the position of $\mu$ is within the band
regime, RTP (TATP) dominates the electron transport between the
electrodes at low (high) temperatures. On the other hand, TATP
fully dominates the electron transport, when the position of $\mu$
is outside the band regime. In Fig. 3(b) and 3(e), Seebeck
coefficient is highly suppressed in the conducting regime, while
significant $S$ values appear in the insulating regime. We
introduce the picture of hole-transport to describe a positive
Seebeck coefficient. Holes are the empty electronic states below
$\mu$. The maximum $PF$ is given by the configuration shown in
Fig. 2(c) and there exists an asymmetrical electron-hole power
factor, as seen in Fig. 3(c) and 3(f). Since $PF_{max,e}$ is
larger than $PF_{max,h}$, $PF$ prefers the electronic states of
TLs above the chemical potential. We note that the positions of
$\mu$ corresponding to $PF_{max,e}$ values are different at
different temperatures. They are $\mu=-26\Gamma_0$ and
$\mu=-28\Gamma_0$ for $k_BT=2.5\Gamma_0$ and $k_BT=5\Gamma_0$,
respectively. The position of the chemical potential appears at
the left side of LBE for $k_BT=5\Gamma_0$. This indicates that
$PF_{max,e}$ prefers $\mu$ far away from the band center $E_0$ at
high temperatures. Although electron hopping strengths of TLs can
be changed by electric
fields[\onlinecite{GeimAK}-\onlinecite{Desai}], we will only focus
on $t_x=t_c$ configuration in the following discussion.

\begin{figure}[h]
\centering
\includegraphics[angle=-90,scale=0.3]{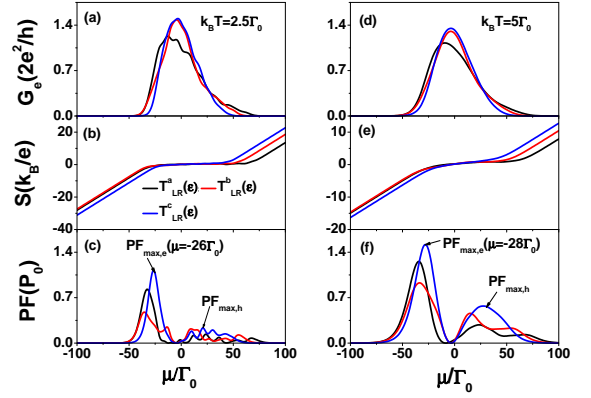}
\caption{(a) Electrical conductance $G_e$, (b) Seebeck coefficient
$S$ and (c) power factor $PF$ as functions of $\mu$ for different
configurations shown in Fig. 2 at $k_BT=2.5\Gamma_0$.
$P_0=\frac{2k^2_B}{h}$. The curves of (d),(e) and (f) are
calculated at $k_BT=5\Gamma_0$. Other physical parameters are the
same as those of Fig. 2. The units of $k_B/e$ and $P_0$ are $86.25
\mu V/K$ and $0.58 pW/K^2$,respectively.}
\end{figure}

To clarify the contact effect between the electrodes and the TLs,
we have calculated $T_{LR}(\varepsilon)$ for different tunneling
rate values ($\Gamma_t$) at $t_x=t_c=6\Gamma_0$ in Fig. 4(a).
$T_{LR}(\varepsilon)$ is distributed between $LBE=-18\Gamma_0$ and
$UBE=36~\Gamma_0$. A large enhancement of $T_{LR}(\varepsilon)$ is
observed as $\Gamma_t$ increases. Nevertheless, such an
enhancement exists only for positive $\varepsilon$ when we further
increases tunneling rate up to $\Gamma_t=12\Gamma_0$. To examine
contact effect on thermoelectric coefficients, we have calculated
$G_e$, $S$ and $PF$ in Fig. 4(b)-4(d) as functions of $\mu$ at
$k_BT=2.5\Gamma_0$. The maximum $G_e$ occurs at $\mu=-5,-1$ and
$2\Gamma_0$ for $\Gamma_t=1,6$ and $12\Gamma_0$, respectively. $S$
is vanishingly small when $G_e$ reaches a maximum value. Unlike
$G_e$, $S$ is not sensitive to the variation of $\Gamma_t$. As a
consequence, the trend of power factor with respect to $\Gamma_t$
is same as the trend of $G_e$. However, $PF_{max,e}$ occurs at
$\Gamma_t=6\Gamma_0$ but not $\Gamma_t=12\Gamma_0$. On the other
hand, $PF_{max,h}$ occurs at $\Gamma_t=12\Gamma_0$. It is worth
noting that $PF_{max,e}$ ($PF_{max,h}$) is given by
$\mu=-16\Gamma_0$ ($\mu=33\Gamma_0$),), which approaches the LBE
(UBE). The results in Fig. 4 show that the optimization of PF
largely depends on the contact properties.

\begin{figure}[h]
\centering
\includegraphics[angle=-90,scale=0.3]{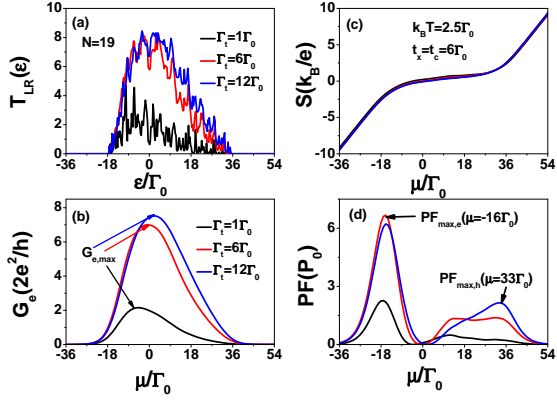}
\caption{(a) Transmission coefficient as functions of
$\varepsilon$ for various tunneling rates at $t_x=t_c=6~\Gamma_0$
and $N=19$. (b) Electrical conductance $G_e$, (c) Seebeck
coefficient $S$ and (d) power factor $PF$ as functions of $\mu$
for different tunneling rates ($\Gamma_t$) at $k_BT=2.5\Gamma_0$,
$t_x=t_c=6\Gamma_0$ and $N=19$.}
\end{figure}

To reveal the size effect of TLs, we have calculated $G_e$, $S$,
and $PF$ in Fig. 5 as functions of $\mu$ for different $N$ values
at $k_BT=2.5\Gamma_0$. As seen in Fig. 5(a), $G_e$ increases with
increasing $N$. It is attributed to the increasing area of
$T_{LR}(\varepsilon)$. The enhancement of $G_e$ will unavoidably
suppress $S$ because $S$ is related to $G_e$. Nevertheless, we see
that the enhancement of $G_e$ with increasing electronic states
does not suppress $S$ in Fig. 5(b). As a consequence, we observe
the enhancement of $PF$ with increasing $N$ in Fig. 5(c). A
remarkable thermoelectric device needs a high efficiency and
significant power output. Now we discuss the dimensionless figure
of merit $ZT$, which is given by

\begin{eqnarray}
ZT &=&\frac{S^2G_eT}{\kappa_e+\kappa_{ph}}\\ \nonumber
&=&\frac{\Lambda}{1-\Lambda+A},
\end{eqnarray}
where $\Lambda={\cal L}^2_1/({\cal L}_0{\cal L}_2)$, and
$A=\frac{T\kappa_{ph}}{{\cal L}_2}$. $\kappa_{ph}$ is phonon
thermal conductance. One can find that the largest value of $ZT$
is given by $\Lambda \rightarrow 1$ and $A \rightarrow 0$. Fig.
5(d) shows the maximum $ZT_{max}=S^2/L_0$ value as functions of
$\mu$ at $k_BT=2.5\Gamma_0$ in the case of $\kappa_{ph}=0$
($A=0$). $L_0=\kappa_e/(TG_e)$ is the Lorenz number. The maximum
$ZT$ values are $ZT_{LBE}=3.3$ and $ZT_{UBE}=2.9$ for the
positions of $\mu$ at LBE and UBE, respectively. We note that
$ZT_{LBE}$ is greater than three. Although $(ZT)_{max}$ shows a
significant value when $\mu$ is far away from the $LBE$, its $PF$
becomes vanishingly small. When compared with the band structures
of TMDC and TMO materials calculated by the first-principle
method,[\onlinecite{Fan}-\onlinecite{Moon}] we estimate $\Gamma_0$
to be between $5~meV$ and $15~meV$. In this work, we adopte
$\Gamma_0=10~meV$. The maximum $PF$ of electrons (holes) at room
temperature ($k_BT=25~meV$) occurs at near the LBE (UBE).

\begin{figure}[h]
\centering
\includegraphics[angle=-90,scale=0.3]{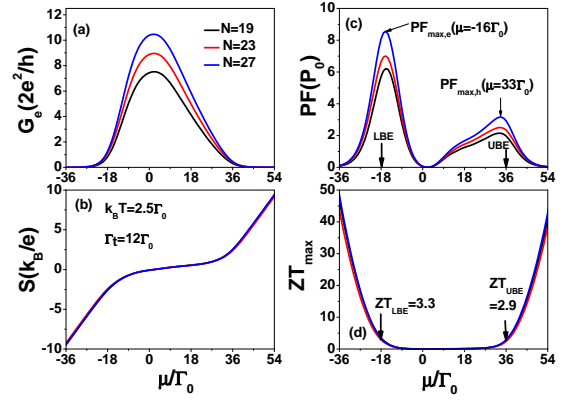}
\caption{(a) Electrical conductance, (b) Seeback coefficient, (c)
power factor and (d) figure of merit ($ZT$) as functions of $\mu$
for various $N$ values at $\Gamma_t=12~\Gamma_0$ and
$k_BT=2.5~\Gamma_0$. Other physical parameters are the same as
those of Fig.4.}
\end{figure}
In the previous results, we have focused on a TL with $N_x=N_y$
condition. Fig. 6 shows the calculated $G_e$, $S$ and $PF$ as
functions of $\mu$ for different $N_y$ values at
$k_BT=2.5\Gamma_0$ and $N_x=51$. $G_e$ increases with increasing
$N_y$. Nevertheless, $G_e$ does not change if one increases $N_x$
($N_x> 51$) at a fixed $N_y$ value (not shown here). This implies
that the enhancement of $G_e$ with increasing electronic states is
attributed to increasing the contact area between the electrodes
and the TL. As shown in Fig. 6(b), $S$ does not change when $G_e$
increases with increasing electronic states. As seen in Fig. 6(c),
the maximum  $PF$ of electrons occurs near LBE even though the TL
shows a nanoribbon pattern. Because the energy harvesting of
thermoelectric devices is expected to operate in a wide
temperature range, we calculate $G_e$, $S$ and $PF$ as functions
of temperature for different $\mu$ values at $N_y=7$ and $N_x=51$
in diagrams (d), (e) and (f), in that order. For
$\mu=-16\Gamma_0$, a finite $G_e$ is mainly contributed from RTP
at low temperatures. When $\mu=-20\Gamma_0$, the electron
transport is dominated by the TATP. As a result, $G_e$ is
numerically significant only at high temperatures. $S$ is very
sensitive to $\mu$ as $T \rightarrow 0$. The curve of
$\mu=-16\Gamma_0$ shows the best $PF$ in a wild temperature range
($k_BT \le 5\Gamma_0$), as seen in Fig. 6(d).

\begin{figure}[h]
\centering
\includegraphics[angle=-90,scale=0.3]{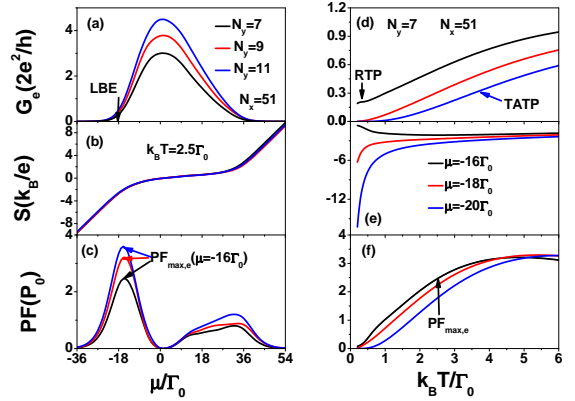}
\caption{(a) Electrical conductance $G_e$, (b) Seeback coefficient
$S$, and (c) power factor $PF$ as functions of $\mu$ for various
$N_y$ values at $N_x=51$ and $k_BT=2.5~\Gamma_0$. Other physical
parameters are the same as those of Fig. 5. (d), (e) and (f) are,
respectively, $G_e$, $S$ and $PF$ as functions of temperature for
different $\mu$ values at $N_y=7$ and $N_x=51$.}
\end{figure}
So far, our discussions are restricted to the one-band model. The
proximity effect between the bands that influences the
thermoelectric properties of ITDMs should also be clarified. To
address this problem, we consider that each band has homogenous
electronic states and that $T_{LR}(\varepsilon)$ is given by

\begin{equation}
T_{LR}(\varepsilon)=\left\{ \begin{array}{ll}
A_c& \mbox {if $E_{CBM}+\Delta_c \ge \varepsilon \ge E_{CBM},$}\\
A_v& \mbox {if $E_{VBM}\ge \varepsilon \ge E_{VBM}+\Delta_v,$}\\
0&\mbox{otherwise}\end{array} \right.
\end{equation}
$E_{CBM}$ and $E_{VBM}$ denote the conduction band minimum and
valence band maximum, respectively. Their band widths are
$\Delta_c$ and $\Delta_v$. Band gap is defined as
$E_g=E_{CBM}+|E_{VBM}|$. Although Eq. (8) is too simple to
describe the phenomena resulting from the variances in size,
electron hopping strength, contact, and geometry,we can obtain a
closed-form solution of the Seebeck coefficient. The analytical
forms of $G_e=G_{e,c}+G_{e,v}$ and $S=(S_c+S_v)/G_e$ are
\begin{equation}
G_{e,c}=-\frac{e^2A_c}{2h}~(tanh(y_1)-tanh(y_2)),
\end{equation}

\begin{equation}
G_{e,v}=-\frac{e^2A_v}{2h}~(tanh(z_1)-tanh(z_2)),
\end{equation}

\begin{eqnarray}
S_c&=&\frac{ek_BA_c}{h}[(y_1~tanh(y_1)-log(cosh(y_1)))\\ \nonumber
&-&(y_2~tanh(y_2)-log(cosh(y_2)))]
\end{eqnarray}
and

\begin{eqnarray}
S_v&=&\frac{ek_BA_v}{h}[(z_1~tanh(z_1)-log(cosh(z_1)))\\ \nonumber
&-&(z_2~tanh(z_2)-log(cosh(z_2)))]
\end{eqnarray}
with variables  $y_1=\frac{E_{CBM}-\mu}{2k_BT}$,
$y_2=\frac{E_{CBM}+\Delta_c-\mu}{2k_BT}$,$z_1=\frac{E_{VBM}+\Delta_v-\mu}{2k_BT}$,
and $z_2=\frac{E_{VBM}-\mu}{2k_BT}$. Using Eqs. (9)-(12), we have
calculated $S$ and $PF$ as functions of $\mu$ for two different
temperatures in Fig. 7. To reveal the proximity effect, two curves
considering one-band model ($A_v=0$) are also plotted. The second
band dramatically changes the behavior of $S$, which is
vanishingly small near the center of the band gap. The proximity
effect can be ignored if the ratio of $E_g/(2k_BT)$ is greater
than ten. This indicates that the prediction of thermoelectric
properties in a one-band model is valid as long as $E_g/(2k_BT)
\ge 10$. Finally, we ask how electron Coulomb interactions
influence the thermoelectric properties of a TL. If the wave
functions of the electrons in each lattice are localized, the
electron Coulomb interactions are strong. Their effects on
electron transport are significant in the scenario of weak hopping
strengths.[\onlinecite{Kuo2}] On the other hand, the wave
functions of the electrons are delocalized in the scenario of
strong hopping strengths to form bands; hence their weak electron
Coulomb interactions can be ignored. Our study belongs to the
latter case.

\begin{figure}[h]
\centering
\includegraphics[angle=-90,scale=0.3]{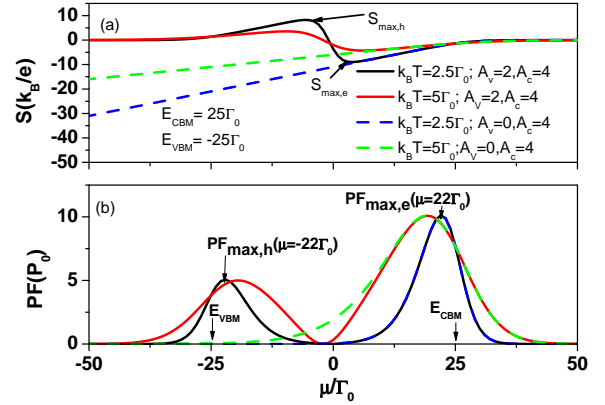}
\caption{(a) Seebeck coefficient $S$ and (b) power factor $PF$ as
functions of $\mu$ for two different temperatures at
$\Delta_c=-\Delta_v=54~\Gamma_0$, $E_{CBM}=25~\Gamma_0$ and
$E_{VBM}=-25~\Gamma_0$. Two curves considering one band model
($A_v=0$) are also plotted.}
\end{figure}

\section{Conclusion}
We theoretically studied the thermoelectric properties of finite
2D TLs coupled to electrodes based on the framework of the
tight-binding model, which does not need heavily numerical
calculations like the first-principle methods. Electron-hole
symmetry breaking appears in the power factor ($PF$). This is
attributed to the $T_{LR}(\varepsilon)$ without the inversion
symmetry of $\varepsilon$. In a negative S regime, a steep change
in the transmission coefficient gives rise to a large electrical
conductance. This explains why the maximum $PF$ of electrons is
larger than that of holes. According to Fig. 4, the contact
between the electrodes and a 2D TL significantly influences the
optimization of $PF$. In particular, the peak of $PF$ for
electrons at room temperature occurs at the position of chemical
potential near the LBE of the 2D TLs. Besides, the enhancement of
Ge with increasing electronic states will not suppress S. Such a
remarkable thermoelectric property is very useful for achieving
large $PF$ values. Finally, we have demonstrated that the one-band
model prediction is adequate when the band gap is ten times larger
than $k_B T$.

%\begin{flushleft}

%\end{flushleft}

%\begin{flushleft}
{\bf Acknowledgments}\\
We are grateful to Dr. Yia-Chung Chang for many encouraging
conversations.
%\end{flushleft}

\mbox{}\\
E-mail address: mtkuo@ee.ncu.edu.tw\\
%E-mail address: yiachang@gate.sinica.edu.tw\\

%\renewcommand{\thesection}{\mbox{Appendix s}} %\section{Appendix}~\Roman{section}
\setcounter{section}{0}

\renewcommand{\theequation}{\mbox{A.\arabic{equation}}} %\section{Appendix}
\setcounter{equation}{0} % reset counter

%\section{}
%\subsection{Derivation of the tunneling current formula using Dyson's equations\label{App:TC_l} }
\mbox{}\\
%{\bf Appendix A. }

%{\bf Data Availability Statements}\\

%The data that supports the findings of this study are available
%within the article.


\begin{thebibliography}{100}

%\begin{itemize}
%\begin{enumerate}

\bibitem[1]{Mahan} G. D. Mahan and J. O. Sofo, Proc. Natl. Acad.
Sci. USA 93, 7436 (1996).

\bibitem[2]{MahanG} G. D. Mahan, L. M. Woods, Phys. Rev. Lett.
\textbf{80}, 4016 (1998).

\bibitem[3]{Chen} G. Chen, Phys. Rev. B \textbf{57},
14958 (1998).

\bibitem[4]{ChenG} G. Chen, M. S. Dresselhaus, G. Dresselhaus, J. P.
Fleurial, and T. Caillat, International Materials Reviews,
\textbf{48}, 45 (2003).

\bibitem[5]{Kuo1} David. M.-T. Kuo and Y. C. Chang, Phys. Rev. B \textbf{81},
205321 (2010).

\bibitem[6]{Whitney1} R. S. Whitney, Phys. Rev. Lett. \textbf{112}, 130601
(2014).

\bibitem[7]{Kuo2} David. M. T. Kuo, C. C. Chen and Y. C. Chang, Phys. Rev. B
\textbf{95}, 075432 (2017).


\bibitem[8]{Venkatasubramanian} R. Venkatasubramanian, E. Siivola,T. Colpitts,B. O¡¦Quinn,
Nature \textbf{413}, 597 (2001).

\bibitem[9]{ Boukai} A. I. Boukai, Y. Bunimovich, J. Tahir-Kheli, J. K. Yu, W. A.
Goddard III and J. R. Heath, Nature, \textbf{451}, 168 (2008).

\bibitem[10]{Harman} T. C. Harman, P. J. Taylor, M. P. Walsh, B. E. LaForge,
Science \textbf{297}, 2229 (2002).

\bibitem[11]{Kuo3} David. M. T. Kuo and Y. C. Chang,
Nanotechnology,\textbf{ 24}, 175403 (2013).


\bibitem[12]{Kagan} T. C. Kagan and C. B. Murry, Nat.
Nanotechnology\textbf{ 10}, 1013 (2015).


\bibitem[13]{GeimAK} A. K. Geim, and I. V. Grigorieva, Nature \textbf{499},
419 (2013).

%"Van der Waals heterostructures"

\bibitem[14]{Novoselov} K. S. Novoselov, A. Mishchenko, A. Carvalho, and A. H. C. Neto,
Science \textbf{353}, aac9439 (2016).

%"2D materials and van der Waals heterostructures"

\bibitem[15]{Desai} S. B. Desai et al., Nano Lett. \textbf{14}, 4592 (2014).


\bibitem[16]{Zhao} L. D. Zhao, S. H. Lo, Y. Zhang, H. Sun, G. Tan,
C. Uher, C. Wolverton, V. P. Dravid and M. G. Kanatzidis, Nature
\textbf{508}, 373 (2014).

%"Ultralow thermal conductivity and high thermoelectric figure of
%merit in SnSe crystals"

%\bibitem[17]{ChenY} Y. Chen et al, Advanced Materials 31, 1804979
%(2019).

%"In-Plane Anisotropic Thermal Conductivity of Few-Layered
%Transition Metal Dichalcogenide Td-WTe2"


\bibitem[17]{Hippalgaonkar} K. Hippalgaonkar, Y. Wang, Y. Ye, D. Y. Qiu,
H. Zhu,Y.  Wang, J. Moore, S. G. Louie and  X. Zhang, Phys. Rev. B
\textbf{95}, 115407 (2017).

%High thermoelectric power factor in two-dimensional crystals of
%MoS2


\bibitem[18]{ChangC} C. Chang, M. Mu, D. He, Y. Pei, and C. F. Wu et
al., Science \textbf{360}, 778 (2018).

%"3D charge and 2D phonon transports leading to high out-of-plane
%ZT in n-type SnSe crystals"



%\bibitem[16]{Chosh} S. Chosh et al. Appl. Phys. Lett. \textbf{92}, 151911
%(2008). graphene


\bibitem[19]{Fan} D. D. Fan, H. J. Liu, L. Cheng, P. H. Jiang, J. Shi and X. F. Tang,
Appl. Phys. Lett. \textbf{105}, 133113 (2014).

%"MoS2 nanoribbons as promising thermoelectric materials"


\bibitem[20]{Huang} W. Huang, X. Luo, C. K. Gan, S. Y. Quek and C.
C . Liang, Phys. Chem. Chem. Phys.  \textbf{16}, 10866 (2014).

%"Theoretical study of thermoelectric properties of few-layer MoS2
%and WSe2"

%\bibitem[21]{Pu} J. Pu, K. Kanahashi, N. T. Cuong, C. H. Chen, L.
%J. Li, S. Okada, H. Ohta and T. Takenobu, Phys. Rev. B 94, 014312
%(2016).

\bibitem[21]{Ouyang} Y. L. Ouyang, Y. Xie, Z. W. Zhang, Q. Peng,  and
Y. P. Chen, J. Appl. Phys. \textbf{120}, 235109 (2016).

%"Very high thermoelectric figure of merit found in hybrid
%transition-metal-dichalcogenides"


\bibitem[22]{Ozbal} G. Ozbal, R. T. Senger, C. Sevik and H.
Sevineli, Phys. Rev. B \textbf{100}, 085415 (2019).

%Ballistic thermoelectric properties of monolayer semiconducting
%transition metal dichalcogenides and oxides

\bibitem[23]{Moon} H. Moon, J. Bang, S. Hong, G. Kim, J. W.
Roh, J. Kim, and W. Lee, ACS Nano,\textbf{ 13}, 13317 (2019).
%"Strong Thermopower Enhancement and Tunable Power Factor via
%Semimetal to Semiconductor Transition in a Transition-Metal
%Dichalcogenide"

\bibitem[24]{Wang} Y. Wang et al, Nature \textbf{568}, 70 (2019).
%"Van der Waals contacts between three-dimensional metals and two-dimensional
%semiconductors"

\bibitem[25]{Qian} X. F. Qian, J. W. Liu,  L. Fu, and J. Li,
Science \textbf{346}, 1344 (2014).

%"Quantum spin Hall effect in two-dimensional transition metal
%dichalcogenides"

\bibitem[26]{WangXM} X. M. Wang et al, Nat. Nanotechnology \textbf{10},
517 (2015).

%"Highly anisotropic and robust excitons in monolayer black
%phosphorus"

\bibitem[27]{WangY} Y. Wang et al, Nature \textbf{550}, 487 (2017).
%"Structural phase transition in monolayer MoTe2 driven by electrostatic
%doping"

\bibitem[28]{LuAY} A. Y. Lu et al, Nat. Nanotechnology \textbf{12}, 744
(2017).

%"Janus monolayers of transition metal dichalcogenides"


\bibitem[29]{Li} M. Y. Li et al, Science \textbf{349}, 524 (2015).

%"Epitaxial growth of a monolayer WSe2-MoS2 lateral p-n junction
%with an atomically sharp interface"


\bibitem[30]{Franklin} A. D. Franklin, Science \textbf{349}, aab2750
(2015).

%"Nanomaterials in transistors: From high-performance to thin-film
%applications"



\bibitem[31]{Akinwande} D. Akinwande et al, Nature \textbf{573}, 507
(2019).

%"Graphene and two-dimensional materials for silicon technology"


\bibitem[32]{LiJ} J. Li et al, Nature \textbf{579}, 368 (2020).

%"General synthesis of two-dimensional van der Waals
%heterostructure arrays"


%\bibitem[29]{ZhangG} G. Zhang and Y. W. Zhang, J. MATERIALS CHEMISTRY
%C 5, 7684 (2017).


\bibitem[33]{Haug} H. Haug and A. P. Jauho, Quantum Kinetics in Transport
and Optics of Semiconductors (Springer, Heidelberg, 1996).


%\bibitem[31]{Meir} Y. Meir and N. S. Wingreen, Phys. Rev. Lett. \textbf{68},
%2512 (1992).

\bibitem[34]{Kuo4} D. M. T. Kuo, AIP Advances \textbf{10}, 045222 (2020).

\bibitem[35]{Horiguchi} T. Horiguchi, Physica A \textbf{178}, 351 (1991).




%\bibitem[3]{Kuo2} David M. T. Kuo and Y. C. Chang, Physica E
%\textbf{115}, 113671 (2020).
















\end{thebibliography}
\end{document}